\documentclass[final,3p,times]{elsarticle}
\usepackage{amssymb,bbold}
\usepackage{amsmath}
\usepackage[utf8x]{inputenc}
\usepackage{natbib}
\usepackage{hyperref}

\hypersetup{
  colorlinks=true,linkcolor=blue,citecolor=blue,
  filecolor=blue,urlcolor=blue,breaklinks=true
}

\biboptions{sort&compress}
\journal{Physica A}
\RequirePackage{color}

\newcommand{\orcid}[1]{\href{https://orcid.org/#1}
{\includegraphics[width=7pt]{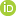}}}

\begin{document}

\begin{frontmatter}

\title{Enhancement of photon creation through the pseudo-Hermitian Dynamical Casimir Effect}

\author[ppg]{D. Cius\orcid{0000-0002-4177-1237}}
\ead{danilocius@gmail.com}

\author[ppg,uepg]{F. M. Andrade\orcid{0000-0001-5383-6168}}
\ead{fmandrade@uepg.br}

\author[uepgdf]{A. S. M. de Castro\orcid{0000-0002-1521-9342}}
\ead{asmcastro@uepg.br}

\author[ifsc]{M. H. Y. Moussa\orcid{0000-0002-3026-0845}}
\ead{miled@ifsc.usp.br}

\address[ppg]{
  Programa de P\'os-Gradua\c{c}\~{a}o Ci\^{e}ncias/F\'{i}sica,
  Universidade Estadual de Ponta Grossa,
  84030-900 Ponta Grossa, Paran\'a, Brazil
}

\address[uepg]{
  Departamento de Matem\'{a}tica e Estat\'{i}stica,
  Universidade Estadual de Ponta Grossa,
  84030-900 Ponta Grossa, Paran\'{a}, Brazil
}

\address[uepgdf]{
  Departamento de F\'{i}sica,
  Universidade Estadual de Ponta Grossa,
  84030-900 Ponta Grossa, Paran\'{a}, Brazil
}

\address[ifsc]{
  Instituto de F\'{\i}sica de S\~{a}o Carlos,
  Universidade de S\~{a}o Paulo,
  Caixa Postal 369,
  13560-970 S\~{a}o Carlos, S\~{a}o Paulo, Brazil
}

\begin{abstract}
We analyse here the pseudo-Hermitian Dynamical Casimir Effect, proposing a non-Hermitian version of the effective Law's Hamiltonian used to describe the phenomenon. 
We verify that the average number of created photons can be substantially increased, a result which calls the attention to the possibility of engineering the time-dependent non-Hermitian Hamiltonian we have assumed. 
Given the well-known difficulty in detecting the Casimir photon production, the present result reinforces the importance of pseudo-Hermitian quantum mechanics as a new chapter of quantum theory and an important tool for the amplification of Hermitian processes such as the degree of squeezing of quantum states.
\end{abstract}


\end{frontmatter}

\section{Introduction}

The Casimir effect \cite{milonni:13,milton:01,woods:16,dodonov:20} refers to the disturbance of the density of vacuum modes caused by uncharged metallic materials placed in close proximity, which generate forces between these boundary surface materials. 
In his seminal work of 1948, H. B. G. Casimir \cite{casimir:48} computed the force $2\pi\hbar c/240d^{4}$ between two parallel metallic plates separated by the distance $d$. 
The Casimir force can be controlled by changing the boundary conditions imposed on the fields by the geometry of the nearby metallic materials \cite{chen:02,bressi:02,krause:07,kim:09,intravaia:13,tang:17,garrett:18} or their optical properties \cite{rehler:71,decca:03,deman:09,banishev:13,somers:18}.

The Dynamical Casimir Effect (DCE), which also follows from the disturbance of the quantum vacuum, refers to particles creation and annihilation now due to accelerating boundaries. 
In the early 1970s, the quantization of the radiation field in a cavity with moving, perfectly reflecting boundaries \cite{moore:70}, contributed to a deepening understanding of the phenomenon. 
The quantum statistical properties of the created photons has been analyzed \cite{sarkar:92,dodonov:90}, showing non-thermal distribution and squeezing properties. 
The thermal effects on the creation of Casimir photons have also being investigated \cite{schaller:02a,schaller:02b,plunien:00}, showing that finite temperatures can enhance the photon number by several orders of magnitude. 
This conclusion is in line with that reached by L. Parker \cite{parker:68,parker:69} in the late 1960s, on the process of particle creation in the expanding universe, where the initial presence of bosons acts to increase the rate of boson creation by the expansion mechanism. Interestingly, the situation is reversed for fermions.

The formulation of an effective Hamiltonian for the DCE by C. K. Law \cite{law:94,law:95}, enabled the derivation of the essential features of the phenomenon through the Schr\"{o}dinger equation. 
This effective Hamiltonian also made it possible to derive the explicit form of the cavity field state and its characteristic resonances with the oscillatory boundaries. 
The quadratic structure of this effective Hamiltonian accounts for the instantaneous modes of the cavity, parametrically amplified and coupled to each other due to the moving boundaries. 
When considering only the cavity mode under the parametric resonance condition, the time-dependent (TD) Law's effective Hamiltonian reads
\begin{equation}
H(t)=\omega(t)a^{\dag}a-i\zeta(t)\left(  a^{2}-a^{\dag}{}^{2}\right)  ,
\label{1}
\end{equation}
where $a^{\dag}$ and $a$ are the creation and annihilation operators of the cavity mode which, in general, is a rapidly modulated frequency $\omega(t)=\omega_{0}[1+\varepsilon\cos(\kappa t)]$, with a small modulation depth $\varepsilon \ll 1$, $\omega_{0}$ being the natural frequency of the mode and $\kappa$ the frequency of oscillation of the moving boundary. 
The strength of the parametric amplification process is given by $\zeta(t)=\left[  4 \omega(t)\right]^{-1}d\omega(t)/dt \approx - \left(\varepsilon\kappa/2\right)  \sin(\kappa t)$.

The Law's effective Hamiltonian has been used for the continuous monitoring of the DCE with a damped detector \cite{decastro:14}, and a dissipative counterpart of this Hamiltonian has also been derived \cite{celeri:09}, leading to a comprehensive analysis of the decoherence mechanisms within the nonideal DCE. 
The Hamiltonian presented in Ref. \cite{celeri:09} has also been used to compute a general expression for the average number of created particles (which applies to any law of motion for the boundary, under the only restriction of small velocities), enabling the draw of the whole scenario of the emerging resonances. 
An analogue model for DCE has been proposed for cavity optomechanical systems \cite{motazedifard:15}, in which the periodic modulation of the cavity length is shown to be equivalent to a time-dependent modulation of the pump field in an optical parametric amplifier into the cavity.

The exceedingly small rate of photon creation in the DCE, coming from the parametric resonance in Hamiltonian (\ref{1}), has challenged experimental measurements, thus leading to a continuous improvement of the experimental proposals towards material science \cite{dodonov:09}.  
Experimental schemes have been theoretically suggested to DCE generation of photons and phonons excitation in optomechanical systems in dispersive regime \cite{motazedifard:17}, and in hybrid cavity interacting with Bose–Einstein condensate under quantum noise and dissipation \cite{motazedifard:18}.

The aim of the present contribution is to consider a pseudo-Hermitian effective Hamiltonian for the DCE, seeking for a significant increasing of the rate of Casimir's photon creation.
Our expectations are supported by recent works on TD pseudo-Hermitian quadratic bosonic Hamiltonians \cite{deponte:19,dourado:20,dourado:21}, which demonstrate the possibility of reaching an infinite degree of squeezing at a finite time interval. 
This follows, essentially, from the fact that pseudo-Hermitian Hamiltonians enables much higher pumping rates that those of Hermitian processes, enough to allow an infinite transfer of energy in finite times; or at least allows this energy transfer trend, described by a finite-time diverging exponential, to occur for a given period of time, what is in fact good enough. 
Both processes, of TD cavity-field squeezing and Casimir's photon production, are described by TD quadratic bosonic Hamiltonians that are very similar to each other, and a pseudo-Hermitian effective Hamiltonian ruling the DCE thus support our expectations of achieving a strong increase in the photon production rate as occurs with the squeezing rate. 

An effective quadratic Hamiltonian for the DCE is of the form
\begin{equation}
H(t)=\omega(t)\left(  a^{\dag}a+1/2\right)  +\alpha(t)a^{2}+\beta(t)a^{\dag}{}^{2}, 
\label{2}
\end{equation}
with the TD complex coefficients of the unbalanced parametric amplification defined as $\alpha(t)=-i\tilde{\alpha}(t)\zeta(t)$ and $\beta(t)=i\tilde {\beta}(t)\zeta(t)$, with real $\tilde{\alpha}(t), \tilde{\beta}(t)\in [0,1]$ and the real frequencies $\omega(t)$ and $\zeta(t)$ being defined above. 
The temporal dependencies of the dimensionless coefficients $\tilde{\alpha}(t)$ and $\tilde{\beta}(t)$ are considered here only for the sake of generality; to favor either \textit{the possibility of emulating a real non-Hermitian DCE, described by the Law's-like Hamiltonian} (\ref{2}), or \textit{the possibility of simulating the pseudo-Hermitian DCE through an engineered effective TD non-Hermitian Hamiltonian} (\ref{2}). 
In either case, the essential connection with Law's Hamiltonian (\ref{1}) lies in the relation of $\alpha(t)$ and $\beta(t)$ with the strength $\zeta(t)$.

Although we must consider the time-dependence of the parameters $\tilde{\alpha}(t)$ and $\tilde{\beta}(t)$ for all the calculations throughout the work, for the final derivation of the mean rate of photon creation we consider the simplified forms $\alpha(t)=-i\tilde{\alpha}_{0}\zeta(t)$ and $\beta(t)=i\tilde{\beta}_{0}\zeta(t)$, with $\tilde{\alpha}_{0}$ and $\tilde{\beta}_{0}$ time-independents. 
When demanding the Hamiltonian (\ref{2}) to be $\mathcal{PT}$-symmetric, such that $\mathcal{PT}H(t)\mathcal{PT}=H(-t),$ we must ensure that $\omega(t)=\omega^{\ast}(-t)$, $\alpha(t)=\alpha^{\ast}(-t)$, and $\beta(t)=\beta^{\ast}(-t)$, the first condition being automatically satisfied, while the remaining two demand that $\tilde{\alpha}(t)=\tilde{\alpha}(-t)$ and $\tilde{\beta}(t)=\tilde{\beta}(-t)$, since $\zeta(t)=\zeta^{\ast}(-t)$.

The possibility of engineering such unbalanced effective Hamiltonian (\ref{2}) for the DCE remains to be investigated. 
The engineering of interactions played, in the last decades, a prominent role in the field of radiation-matter interaction, focused on the construction of specific quantum states \cite{villas:05,villas:03a,villas:03b,villas:01,serra:00,parkins:93,parkins:95,garraway:94,law:96,vogel:93,moussa:98} and reservoirs \cite{villas:03,dealmeida:04,serra:05,prado:06,neto:12,neto:13,prado:13,rosado:15,rossetti:16} through the method of adiabatic elimination of fast variables \cite{gamel:10,james:07}. 
The developments here undertaken may certainly be applied to the construction of effective autonomous or nonautonomous pseudo-Hermitian Hamiltonians.

The recent field of pseudo-Hermitian quantum mechanics began in 1998 with the work by Bender \& Boettcher \cite{bender:98} demonstrating that autonomous Hamiltonian which is invariant under parity-time ($\mathcal{PT}$)
transformation, i.e., $[H,\mathcal{PT}]=0$, can exhibit an entirely real spectrum. 
When in addition to being invariant, the Hamiltonian shares its eigenstates with the $\mathcal{PT}$ operator, it is said that it preserves the $\mathcal{PT}$ symmetry, which is broken when $H$ and $\mathcal{PT}$ stop sharing the same eigenstates \cite{bender:07}. 
In 2002, Mostafazadeh \cite{mostafazadeh:02a,mostafazadeh:02b,mostafazadeh:02c} turned to the one remaining problem associated with a non-Hermitian $\mathcal{PT}$-symmetric autonomous Hamiltonian: a probabilistic interpretation of the associated quantum theory. 
Such an indefiniteness of the unitarity of time evolution was overcome by constructing a Hilbert space based on a time independent metric operator where the non-Hermitian Hamiltonian enjoys self-adjointness. 
When a TD metric operator is considered for treating TD non-Hermitian Hamiltonians, different proposals have been presented for reestablishing the self-adjointness to these TD Hamiltonians \cite{znojil:08,gong:13,fring:16a,luiz:20}. 
Many interesting contributions have been presented regarding TD $\mathcal{PT}$-symmetric Hamiltonians \cite{fring:17,maamache:17,fring:19,khantoul:17,mana:20,koussa:20}.

Our manuscript is organized as follows. In Sec. II we present a brief review of the treatment of TD non-Hermitian Hamiltonians with TD metric operators. 
In Sec. III we define a Dyson map to derive the Hermitian counterpart of the effective non-Hermitian Hamiltonian for the DCE. 
The solutions of the Schr\"{o}dinger equations for the TD non-Hermitian Hamiltonian and its Hermitian counterpart, derived from the Lewis \& Riesenfeld method \cite{lewis:69}, are presented in Sec. IV. 
The expression for the average number of photon creation is computed in Sec. V and in Sec. VI we compute the parameters defining the Dyson map and the cavity-mode squeezing, which result in the enhancement of the rate of Casimir's photon creation. 
Finally, in Sec. VII we discuss the experimental implementation of our proposal.

\section{TD non-Hermitian Hamiltonians and metric operators}

In this section we revisit the approach usually considered for describing time dependent non-Hermitian quantum system. We start reviewing the application of a TD metric operator for treating a TD non-Hermitian Hamiltonian by following the proposal in Ref. \cite{fring:16a}. Considering the TD Dyson map $\eta(t)$, we transform the Schr\"{o}dinger equation for the non-Hermitian $H(t)$ written as $i\partial_{t}\left\vert \Psi(t)\right\rangle =H(t)\left\vert \Psi(t)\right\rangle $ into the equation in the form $i\partial_{t}\left\vert \psi(t)\right\rangle =h(t)\left\vert \psi(t)\right\rangle $ $\left(  \hbar=1\right)  $, with the vector states related as $\left\vert \psi(t)\right\rangle  = \eta(t)\left\vert\Psi(t)\right\rangle $, and the Hamiltonians satisfying 
\begin{equation}
h(t)=\eta(t)H(t)\eta^{-1}(t)+i\left[  \partial_{t}\eta(t)\right]  \eta^{-1}(t). 
\label{3}
\end{equation}
By imposing the Hamiltonian $h(t)$ to be a Hermitian operator, the TD quasi-Hermiticity relation is obtained as
\begin{equation}
H^{\dag}(t)\Theta(t)-\Theta^{\dag}(t)H(t)=i\partial_{t}\Theta(t), 
\label{4}
\end{equation}
where $\Theta(t)=\eta^{\dag}(t)\eta(t)$ is the so-called TD metric operator, which ensures the TD probability densities in the Hermitian and non-Hermitian systems to be related as
\begin{equation}
\left\langle \Psi(t)\left\vert \tilde{\Psi}(t)\right. \right\rangle_{\Theta(t)}=\left\langle \Psi(t)\left\vert \Theta(t)\right\vert \tilde{\Psi}(t)\right\rangle =\left\langle \psi(t)\left\vert \tilde{\psi}(t)\right.
\right\rangle .
\label{5}
\end{equation}
Furthermore, the observables $o(t)$ and $O(t)$ in the Hermitian and non-Hermitian systems are tied by means of
\begin{equation}
O(t)=\eta^{-1}(t)o(t)\eta(t), 
\label{6}
\end{equation}
which implies on the equalities
\begin{equation}
\left\langle \Psi(t)\left\vert O(t)\right\vert \tilde{\Psi}(t)\right\rangle_{\Theta(t)}=\left\langle \Psi(t)\left\vert O(t)\Theta(t)\right\vert
\tilde{\Psi}(t)\right\rangle =\left\langle \psi(t)\left\vert o(t)\right\vert
\tilde{\psi}(t)\right\rangle . 
\label{ME}
\end{equation}

Assuming time-independent Dyson map $\eta$ with the metric operator as $\Theta=\eta^{\dag}\eta$, Eqs. (\ref{3}) and (\ref{4}) reduce to the forms
\begin{subequations}
\label{7}
\begin{align}
h(t)  &  =\eta H(t)\eta^{-1},
\label{7a}
\\
H^{\dag}(t)\Theta &  =\Theta H(t). 
\label{7b}
\end{align}
\end{subequations}
Moreover, the relation between the operators $O(t)$ and $o(t)$ according to the Eq. (\ref{6}) considered now for time-independent  $\eta$ 
ensures the pseudo-Hamiltonian $H(t)=\eta^{-1}h(t)\eta$ to be an observable of the system differently from what happens with the case of a TD Dyson map, where $H(t)$ and its Hermitian counterpart $h(t)$ are not related by a similarity transformation. 
Therefore, the price for TD Dyson map and metric operators --- which can prevent unwanted constraints on the TD parameters of $H(t)$ when the Hermiticity is imposed through Eq. (\ref{4}) --- is a nonobservable non-Hermitian Hamiltonian $H(t)$. 
Although TD Dyson map and metric operators prevents $H(t)$ from being an observable, the same is not true for all other observables of the system described by $O(t)$ satisfying
Eq. (\ref{6}).

\subsection{The Hermitian counterpart of the effective non-Hermitian Hamiltonian for the DCE.}

In order to derive the Hermitian counterpart of our effective non-Hermitian Hamiltonian (\ref{2}), we introduce a TD quadratic Dyson map \cite{musumbu:06,fring:16b,deponte:19,dourado:20,dourado:21} in the form
\begin{equation}
\eta(t)=\exp\left[  \epsilon(t)\left(  a^{\dagger}a+1/2\right)  +\mu
(t)a^{2}+\mu^{\ast}(t)a^{\dagger2}\right]  , 
\label{9}
\end{equation}
which can be rewritten by means of the Gauss decomposition for $SU(1,1)$ as
\begin{equation}
\eta(t)=\exp\left[  \lambda(t)K_{+}\right]  \exp\left[  \ln\Lambda
(t)K_{0}\right]  \exp\left[  \lambda^{\ast}(t)K_{-}\right]  , 
\label{10}
\end{equation}
with the operators $K_{0}=\left(  a^{\dagger}a+1/2\right)  /2$ and $K_{+}=K_{-}^{\dagger}=a^{\dagger2}/2$ satisfying the commutators $\left[  K_{0},K_{\pm}\right]  = \pm K_{\pm}$ and $\left[  K_{+},K_{-}\right]  =-2K_{0}$. 
From now on we omit the time dependence of the variables, and the coefficients in Eq. (\ref{10}) have the forms
\begin{subequations}
\label{11}
\begin{align}
\lambda &  =\frac{2\mu^{\ast}\sinh\Xi}{\Xi\cosh\Xi-\epsilon\sinh\Xi},
\label{11a}
\\
\Lambda &  =\frac{\Xi^{2}}{\left(  \Xi\cosh\Xi-\epsilon\sinh\Xi\right)^{2}}.
\label{11b}
\end{align}
The argument of the hyperbolic functions is defined as $\Xi = \sqrt{\epsilon^{2}-4|\mu|^{2}}$ where $\epsilon$ is a real positive function such that $\epsilon^{2}-4|\mu|^{2}\geq0$. By following Ref. \cite{musumbu:06}, we introduce the free parameter $z=2\mu/\epsilon=\left\vert z\right\vert e^{i\varphi}$ with $|z|\in [0,1]$, and rewrite $\Xi=\epsilon\sqrt{1-\left\vert z\right\vert ^{2}}$. In addition, we have that
\end{subequations}
\begin{equation}
\lambda=\frac{\epsilon|z|}{\Xi}\frac{\tanh{\Xi}}{1-\frac{\epsilon}{\Xi}\tanh{\Xi}}e^{-i\varphi}=\Phi e^{-i\varphi}, 
\label{12}
\end{equation}
which allows us to obtain
\begin{equation}
\epsilon=\frac{1}{2\sqrt{1-|z|^{2}}}\ln\frac{\left(  1+\sqrt{1-\left\vert z\right\vert ^{2}}\right)  \Phi+|z|}{\left(  1-\sqrt{1-\left\vert z\right\vert^{2}}\right)  \Phi+|z|} \,. 
\label{13}
\end{equation}
Defining the real function $\chi$ as
\begin{equation}
\chi=|\lambda|^{2}-\Lambda=\Lambda\frac{4|\mu|^{2}\sinh^{2}\Xi-\Xi^{2}}{\Xi^{2}}=-\frac{2\Phi}{|z|}-1 , 
\label{14}
\end{equation}
we can write the transformation
\begin{equation}
\eta(t)
\begin{pmatrix}
a \,
\\
a^{\dag}
\end{pmatrix}
\eta^{-1}(t)=\frac{1}{\sqrt{\Lambda}}
\begin{pmatrix}
        1      & -\lambda\,
\\
\lambda^{\ast} & -\chi
\end{pmatrix}
\begin{pmatrix}
a \,
\\
a^{\dag}
\end{pmatrix}, 
\label{15}
\end{equation}
which enables us to derive the Hamiltonian
\begin{equation}
h(t)=2W(t)K_{0}+2T(t)K_{-}+2V(t)K_{+}, 
\label{16}
\end{equation}
from Eq. (\ref{3}). The time-dependent coefficients of (\ref{16}) read as
\begin{subequations}
\label{17}
\begin{align}
W(t)  &  = -\frac{1}{\Lambda}\left[  \omega(\Phi^{2}+\chi)+2(\alpha\lambda+\beta\lambda^{\ast}\chi)+i\left(  \lambda\dot{\lambda}^{\ast}-\frac{\dot{\Lambda}}{2}\right)  \right] ,
\label{17a}
\\
T(t)  &  = \frac{1}{\Lambda}\left[  \omega\lambda^{\ast}+\alpha+\beta\lambda^{\ast2}+i\frac{\dot{\lambda}^{\ast}}{2}\right]  , 
\label{17b}
\\
V(t)  &  =\frac{1}{\Lambda}\left[  \omega\lambda\chi+\alpha\lambda^{2} + \beta\chi^{2}+\frac{i}{2}(\dot{\lambda}\Lambda-\dot{\Lambda}\lambda + \dot{\lambda}^{\ast}\lambda^{2})\right] ,
\label{17c}
\end{align}
with dot representing the time derivative. 

As discussed in previous section, the Hamiltonian (\ref{16}) must be Hermitian what implies in the
constraints $W(t)=W^{\ast}(t)$ and $V(t)=T^{\ast}(t)$ which yield the nonlinear system
\end{subequations}
\begin{subequations}
\label{18}
\begin{align}
\dot{|z|}  &  = -|z|^{2}\left\{  \frac{\Phi^{2}+\chi}{\Phi}|\omega|\sin{\varphi_{\omega}} -2\left[|\alpha|\sin{(\varphi-\varphi_{\alpha})}-\chi|\beta|\sin{(\varphi+\varphi_{\beta})}\right]\right\}  +\frac{|z|}{\Phi}\dot{\Phi} ,
\label{18a}
\\
\dot{\Phi}  & = \frac{2}{\chi-1}\left\{  \left(1-\Phi^{2}\right)\left[\Phi|\omega|\sin\varphi_{\omega}-|\alpha|\sin{(\varphi-\varphi_{\alpha})}\right]  -|\beta|\left[(2\chi-1)\Phi^{2}-\chi^{2}\right]\sin{(\varphi+\varphi_{\beta})}\right\} ,
\label{18b}
\\
\dot{\varphi}  &  =2|\omega|\cos\varphi_{\omega}+\frac{2}{(1-\chi)\Phi}\left\{  |\alpha|(1-\Phi^{2})\cos{(\varphi-\varphi_{\alpha})}+|\beta|\left(\Phi^{2}-\chi^{2}\right)\cos{(\varphi+\varphi_{\beta})}\right\}, 
\label{18c}
\end{align}
where we have defined $\alpha=|\alpha|e^{i{\varphi_{\alpha}}}$ and $\beta=|\beta|e^{i{\varphi_{\beta}}}$. 
Note the Eq. (\ref{18a}) may be replaced by the equivalent form
\end{subequations}
\begin{align}
\dot{\Lambda}  &  =-2\Lambda\left\{  \left(  1+\frac{2\Phi^{2}}{\chi-1}\right) |\omega| \sin\varphi_{\omega} -\frac{2\Phi}{\chi-1}\left[  \left\vert \alpha\right\vert \sin\left(  \varphi-\varphi_{\alpha}\right)  -\left\vert \beta\right\vert\left(2\chi-1\right)  \sin(\varphi+\varphi_{\beta})  \right]\right\}  .
\label{19}
\end{align}
Thus, the set of equations (\ref{18}) allows us to rewrite the parameters of (\ref{16}) in a such a way the real frequency $W$ reads as
\begin{equation}
W = |\omega|\cos\varphi_{\omega}+\frac{2\Phi}{\chi-1}\left[  |\alpha|\cos{(\varphi-\varphi_{\alpha})}-|\beta|\cos{(\varphi+\varphi_{\beta})}\right]  . 
\label{20}
\end{equation}
Likewise we can obtain $V=T^{\ast}=|T|e^{-i\varphi_{T}}$ with
\begin{subequations}
\label{21}
\begin{align}
|T|  &  =\frac{1}{|1-\chi|}\left\{|\alpha|^{2}+|\beta|^{2}\chi^{2}-2|\alpha||\beta|\chi\cos(\varphi_{\alpha}+\varphi_{\beta})\right. 
\nonumber\\
&  +\left. \Phi|\omega|\sin{\varphi_{\omega}}\left[\Phi|\omega|\sin{\varphi_{\omega}}-2|\alpha|\sin{(\varphi-\varphi_{\alpha})}+2|\beta|\chi\sin{(\varphi+\varphi_{\beta})}\right]\right\}^{1/2},
\label{21a}
\\
\tan{\varphi_{T}}  &  =-\frac{|\omega|\Phi\cos{\varphi}\sin{\varphi_{\omega}}+|\alpha|\sin{\varphi_{\alpha}} + |\beta|\chi\sin{\varphi_{\beta}}}{|\omega|\Phi\sin{\varphi}\sin{\varphi_{\omega}} - |\alpha|\cos{\varphi_{\alpha}}+|\beta|\chi\cos{\varphi_{\beta}}}\,. 
\label{21b}
\end{align}

As in Refs. \cite{musumbu:06,fring:16b,deponte:19,dourado:20,dourado:21}, we also note the $z$ is the only free parameter determining the Dyson map, with $\Phi$ and ${\varphi}$ coming from the system of coupled equations (\ref{18}), and $\epsilon$ coming from Eq. (\ref{13}). 
This implies that a given pair $\left(  |z|,\Phi\right)  $ must be further corroborated by a real $\epsilon$, such that $|z|>$ $-2\Phi/(1+\Phi^{2})$.

\subsection{Time-evolution of the Hermitian and non-Hermitian systems}

To obtain the solutions of the Schr\"{o}dinger equations for the Hermitian $h(t)$ and non-Hermitian $H(t)$ Hamiltonians, we invoke the Lewis \& Riesenfeld TD invariants method \cite{lewis:69}, and follow the same procedure as described in Refs. \cite{baseia:92,mizrahi:94}. 
Starting from the Schr\"{o}dinger equation $i\partial_{t}|\psi(t)\rangle=h(t)|\psi(t)\rangle$, we can write its formal solution as
\end{subequations}
\begin{equation}
|\psi(t)\rangle=U(t)|\psi(0)\rangle , 
\label{22}
\end{equation}
with the evolution operator in the form
\begin{equation}
U(t)=\Upsilon(t)S[\xi(t)]D[\theta(t)]R[\tilde{\Omega}(t)]S^{\dagger}[\xi(0)].
\label{23}
\end{equation}
Here $S\left[  \xi(t)\right]  =\exp{\left[  \xi(t)\hat{a}^{\dagger2}-\xi^{\ast}(t)\hat{a}^{2}\right]  }$ is the squeezing operator with $\xi(t)=r(t)e^{i\phi(t)}$, being $r(t)$ the squeezing degree and $\phi(t)$ the squeezing direction in phase space \cite{walls:94,scully:97}, which obey the system of differential equations
\begin{subequations}
\label{24}
\begin{align}
\dot{r}(t)  &  = -2|T(t)|\sin{[\varphi_{T}(t)+\phi(t)]},
\label{24a}
\\
\dot{\phi}(t)  &  = -2W(t)-4|T(t)|\coth{2r(t)}\cos{[\varphi_{T}(t)+\phi(t)]} .
\label{24b}
\end{align}
\end{subequations}
Furthermore, $D\left[  \theta(t)\right]  =\exp{\left[  \theta(t)\hat{a}^{\dagger}-\theta^{\ast}(t)\hat{a}\right]  }$ is the displacement operator with $\theta(t)$ satisfying the equation $i\dot{\theta}(t)=\Omega(t)\theta(t)$ where $\Omega(t) = W(t)+2|T(t)|\tanh{r(t)} \cos{[\varphi_{T}(t)+\phi(t)]}$. 
Finally, $\Upsilon(t)=\exp\left[-i\tilde{\Omega}(t)/2\right]  $ is a global phase factor and  $R[\tilde{\Omega}(t)]=\exp\left[-i\tilde{\Omega}(t)a^{\dagger}a\right] $ is a rotation, with $\tilde{\Omega}(t)=\int_{0}^{t}d\tau\Omega(\tau)$. Once the solution of the Schr\"{o}dinger equation for the Hermitian Hamiltonian $h(t)$ is given, the solution for the Hamiltonian $H(t)$ can be immediately obtained as prescribed in Refs. \cite{fring:16b,deponte:19,dourado:20,dourado:21}, and is written as 
\begin{equation}
|\Psi(t)\rangle=\mathcal{U}(t)|\psi(0)\rangle , 
\label{24c}
\end{equation}
with
\begin{equation}
\mathcal{U}(t)=\eta^{-1}(t)U(t)U^{\dag}(0)\eta(0). 
\label{24d}
\end{equation}

Since the main goal is to obtain the amount of created photon due to DCE, the next task consists in determining the mean photon number from the time-evolved quantum state. This can be done by means of the equation $\mathcal{N}(t)=\langle\psi(t)|a^{\dagger}a|\psi(t)\rangle=\langle\psi(0)|U^{\dag}(t)a^{\dagger}aU(t)|\psi(0)\rangle$ by first considering the transformations
\begin{subequations}
\label{25}
\begin{align}
U^{\dag}(t) a U(t)  &  = u(t)a+v(t)a^{\dagger}+w(t) ,
\label{25a}
\\
U^{\dag}(t) a^{\dagger} U(t)  &  =v^{\ast}(t)a+u^{\ast}(t)a^{\dagger}+w^{\ast}(t) , 
\label{25b}
\end{align}
where the TD functions $u(t)$, $v(t)$ and $w(t)$ are given by
\end{subequations}
\begin{subequations}
\label{B}
\begin{align}
u(t)  &  = e^{-i\tilde{\Omega}(t)}\cosh{r(0)}\cosh{r(t)} - e^{i[\tilde{\Omega}(t)+\phi(t)-\phi(0)]}\sinh{r(0)}\sinh{r(t)} , 
\label{Ba}
\\
v(t)  &  = e^{i[\tilde{\Omega}(t)+\phi(t)]}\cosh{r(0)}\sinh{r(t)} - e^{-i[\tilde{\Omega}(t)-\phi(0)]}\sinh{r(0)}\cosh{r(t)} ,
\label{Bb}
\\
w(t)  &  =\theta(0)e^{-i\tilde{\Omega}(t)}\cosh{r(t)} + \theta^{\ast}(0)e^{i[\tilde{\Omega}(t)+\phi(t)]}\sinh{r(t)}, 
\label{Bc}
\end{align}
\end{subequations}
and obey the identity $|u(t)|^{2}-|v(t)|^{2}=1$. From Eqs. (\ref{25}) the mean photon number is written as
\begin{align}
\mathcal{N}(t) 
&= |v(t)|^{2} + |w(t)|^{2}
 + \left[|u(t)|^{2} + |v(t)|^{2}\right]\langle\psi(0)|a^{\dagger}a|\psi(0)\rangle
 + u(t)v^{\ast}(t)\langle\psi(0)|a^{2}|\psi(0)\rangle
 + v(t)u^{\ast}(t)\langle\psi(0)|a^{\dagger2}|\psi(0)\rangle
\nonumber\\
&+ \left[w(t)v^{\ast}(t) + u(t)w^{\ast}(t)\right] \langle\psi(0)|a|\psi(0)\rangle 
 + \left[w(t)u^{\ast}(t) + v(t)w^{\ast}(t)\right] \langle\psi(0)|a^{\dagger}|\psi(0)\rangle, 
\label{Ni}
\end{align}
and assuming initial vacuum state $|\psi(0)\rangle=|0\rangle$ with $r(0)=0$ and $|\theta(0)|=0$, Eq. (\ref{Ni}) becomes
\begin{equation}
\mathcal{N}(t)=\sinh^{2}{r(t)},\label{Nf}
\end{equation}
which is in agreement with the previous values obtained in applying different approaches \cite{decastro:13,dodonov:96,dodonov:20}. 
Once the general aspects of the dynamics is well described, in the next section we discuss how the non-Hermiticity on the Hamiltonian (\ref{2}) can improve the photon generation in DCE.

\section{Enhancement of photon creation in DCE}

\subsection{The derivation of the Dyson map and squeezing parameters}

For our main purpose, we now consider the polar forms of the TD functions defining the unbalanced parametric amplification in the Hamiltonian (\ref{2}) where we have the frequency $\omega(t)=|\omega(t)| = \omega_{0}[ 1 + \varepsilon \cos{(\kappa t)}]$ and the strength of the parametric excitation in (\ref{1}) written as
\begin{equation}
\zeta(t)=\frac{1}{4\omega(t)} \frac{d\omega(t)}{dt} \approx |\zeta(t)| \exp{\left\{ i [  \mathfrak{h}\left[  \sin{(\kappa t)}\right]  \pi+\pi] \right\}}.  
\label{31.1}
\end{equation}
Here $|\zeta(t)|=\left(  \varepsilon \kappa/2\right)  \left\vert \sin{(\kappa t)}\right\vert$ with the Heaviside function
\begin{equation}
\mathfrak{h}[f(t)]= \frac{1 - \mathrm{sgn}[f(t)]}{2} =
\left\{
\begin{array}{ll}
0, & \mathrm{sgn}[f(t)] = +1 \\
1, & \mathrm{sgn}[f(t)] = -1 \\
\end{array} 
\right.,
\label{H}
\end{equation}
and the signal function
\begin{equation}
\mathrm{sgn}[f(t)]= \frac{f(t)}{|f(t)|} =
\left\{
\begin{array}{ll}
-1, & f(t) < 0 \\
+1, & f(t) \geq 0 \\
\end{array} 
\right. .
\label{s}
\end{equation}
Likewise, we have $\alpha(t)=\tilde{\alpha}(t)|\zeta(t)|e^{i\varphi_{\alpha}(t)}$ with $\varphi_{\alpha}(t)=\mathfrak{h}\left[  \sin{(\kappa t)}\right]  \pi + \pi/2$, and $\beta(t)=\tilde{\beta}(t)|\zeta(t)| e^{i\varphi_{\beta}(t)}$ with $\varphi_{\beta}(t)=\mathfrak{h}\left[  \sin{(\kappa t)}\right]  \pi-\pi/2$. 
These polar forms imply to the Hermitian counterpart of Hamiltonian (\ref{2}) the form 
\begin{equation}
h(t)=W(t)\left(  a^{\dagger}a+1/2\right)  +|T(t)|\left(e^{i\varphi_{T}(t)}a^{2}+e^{-i\varphi_{T}(t)}a^{\dagger2}\right) ,
\label{26}
\end{equation}
where
\begin{subequations}
\label{27}
\begin{align}
W  &  = \omega-\frac{2\zeta\Phi}{\chi-1}\left(  \tilde{\alpha}-\tilde{\beta}\right)  \sin{\varphi} ,
\label{27a}
\\
| T |  &  = \left\vert \frac{\zeta\left(  \tilde{\alpha} - \tilde{\beta}\chi\right) }{1-\chi}\right\vert , 
\label{27b}
\\
{\varphi_{T}}  &  = \mathfrak{h}\left[  \sin{(\kappa t)}\right]\pi + \mathfrak{h}(1-\chi)\pi + \mathfrak{h}{(\tilde{\alpha}-\chi\tilde{\beta})\pi+\pi/2} . 
\label{27c}
\end{align}
\end{subequations}
The differential equations in the nonlinear system (\ref{18}) become
\begin{subequations}
\label{28}
\begin{align}
\dot{|z|}  &  =2\zeta|z|^{2}\left(  \tilde{\alpha}-\tilde{\beta}\chi\right)
\cos{\varphi}+\frac{|z|}{\Phi}\dot{\Phi},
\label{28a}
\\
\dot{\Phi}  &  =\frac{2\zeta}{1-\chi}\left\{  \tilde{\alpha}\left(1-\Phi^{2}\right)
+
\tilde{\beta}\left[\left(2\chi-1\right)\Phi^{2}-\chi^{2}\right]\right\}  \cos{\varphi
},
\label{28b}
\\
\dot{\varphi}  &  =2\omega-\frac{2\zeta}{\left(1-\chi\right)\Phi}\left[  \tilde{\alpha
}\left(1-\Phi^{2}\right) + \tilde{\beta}\left(\Phi^{2}-\chi^{2}\right)\right]  \sin{\varphi},
\label{28c}
\end{align}
\end{subequations}
whereas Eq. (\ref{19}) assumes the form
\begin{equation}
\dot{\Lambda}    =\frac{4\zeta\Phi(\Phi^2 - \chi)}{\chi-1}\left[  \tilde{\alpha} - \tilde{\beta}\left(2\chi-1\right)\right]  \cos{\varphi},
\label{28d}   
\end{equation}
where we substituted $\Lambda = \Phi^2 - \chi$ on the right-hand side from Eq. (\ref{19}) in according to the Eq. (\ref{14}).

To solve them, we focus in a particular solution which is obtained by choosing a fixed value for the free parameter module $|z|=2|\mu|/\epsilon\approx1$, close to unity from below since $|z|\in[0,1]$. This means we can write $2|\mu|\approx\epsilon$ and $z\approx e^{i\varphi}$ which lead to the simplified equations

\begin{subequations}
\label{29}
\begin{align}
\dot{\Phi}  &  \approx-2\zeta\Phi\left(  \tilde{\alpha}-\chi\tilde{\beta}\right)  \cos{\varphi}
\nonumber\\
& =\frac{2\zeta}{1-\chi}\left\{  \tilde{\alpha}\left(1-\Phi^{2}\right) + \tilde{\beta}\left[\left(2\chi-1\right)\Phi^{2} - \chi^{2}\right]\right\}  \cos{\varphi} ,
\label{29a}
\\
\dot{\varphi}  & \approx 2\omega - \frac{2\zeta}{\left(1-\chi\right)\Phi}\left[\tilde{\alpha}\left(1-\Phi^{2}\right)+\tilde{\beta}\left(\Phi^{2}-\chi^{2}\right)\right]\sin{\varphi} ,
\label{29b}
\\
\dot{\Lambda}  &  \approx \frac{4\zeta\Phi(\Phi^2-\chi)}{\chi-1}\left[  \tilde{\alpha} - \tilde{\beta}\left(2\chi-1\right)\right]  \cos{\varphi}. 
\label{29c}
\end{align}
\end{subequations}
For $|z|$ close to unity from below, we can consider Eq. (\ref{14}) in approximate form as $\chi\approx-2\Phi-1$, and apply it in Eq. (\ref{29a}). It results in the equation
\begin{equation}
\zeta(\tilde{\alpha}-\tilde{\beta})(\Phi+1)\cos\varphi\approx0, 
\label{30}
\end{equation}
and since $\tilde{\alpha}\neq\tilde{\beta}$, it follows that
\begin{subequations}
\label{31}
\begin{align}
\Phi &  \approx -1 , \label{31a}
\\
\chi &  \approx 1 ,  \label{31b}
\\
\Lambda & \approx 0, \label{31c}
\end{align}
\end{subequations}
with $\Phi$ and $\chi$ bringing closer, respectively, to  $-1$ (from below) and $1$ (from above). 
As a direct consequence of $z\approx e^{i\varphi}$ we have that $\Xi=\epsilon\sqrt{1-\left\vert z\right\vert^{2}}\approx0$ and $\tanh{\Xi/}\Xi\approx1$. 
Moreover, the parameter $\Phi$, defined in Eq. (\ref{12}), becomes $\Phi\approx\epsilon/\left(  1-\epsilon\right)$ which implies in $\epsilon\gg1$ for $\Phi\approx-1$. These results are consistent with the expression $\epsilon\approx\Phi/\left(  1+\Phi\right)  $, following from Eq. (\ref{13}) under the choice $|z|\approx1$.

From the results obtained in Eq. (\ref{31}) we further simplify Eq. (\ref{29b}) for
\begin{align}
\dot{\varphi}  &  \approx2\omega+2\zeta(\tilde{\alpha}+\tilde{\beta})\sin{\varphi} ,
\label{32a}
\end{align}
and considering long time intervals, such that $\omega_{0}t\gg1$, as usually assumed in the DCE analysis,  it follows that
\begin{equation}
\varphi\approx\varphi_{0}+2\omega_{0}t , 
\label{33}
\end{equation}
where the terms with $\varepsilon$ have been neglected.

In what follows, we address the set of equations (\ref{24}) for the squeeze parameters $r$ and $\phi$. 
To simplify these equations, we consider the regime for $r>1$ which implies $\coth{2r\approx1}$, and apply the change of variable
\begin{equation}
\phi^{\prime}=\phi+\pi/2+\mathfrak{h}{(1-\chi)\pi}+\mathfrak{h}{(\tilde{\alpha}-\chi\tilde{\beta})\pi}. 
\label{34}
\end{equation}
The result derived from the condition $\coth{2r\approx1}$ is contrasted with numerical simulations (see Fig. \ref{fig1}). We then obtain from Eqs. (\ref{24}), the following simplified set of differential equations 
\begin{subequations}
\label{35}
\begin{align}
\dot{r}  &  \approx2\zeta\left\vert \frac{\tilde{\alpha}-\chi\tilde{\beta}}{\chi-1}\right\vert \sin{\phi^{\prime}},
\label{35a}
\\
\dot{\phi}^{\prime}  &  \approx-2\omega+\frac{4\Phi\zeta}{\chi-1}(\tilde{\alpha}-\tilde{\beta})\sin{\varphi}+4\zeta\left\vert \frac{\tilde{\alpha}-\chi\tilde{\beta}}{\chi-1}\right\vert \cos{\phi^{\prime}} .
\label{35b}
\end{align}
\end{subequations}
By assuming the constant values $\tilde{\alpha}=\tilde{\alpha}_{0}$ and $\tilde{\beta}=\tilde{\beta}_{0}$, and setting $\varphi_{0}=\pi/2$, it follows that $\phi^{\prime}\approx\phi_{0}^{\prime}-2\omega_{0}t$. 
As a result, the Eq. (\ref{34}) provides 
\begin{equation}
\phi = \phi_{0}^{\prime}-3\pi/2-\mathfrak{h}{(\tilde{\alpha}_{0}-\chi\tilde{\beta}_{0})\pi} -2\omega_{0}t .
\label{phi}
\end{equation}
Moreover, with the choice $\varphi_{0}=\pi/2$, we finally set the Dyson map parameters $\epsilon$ and $\mu=|\mu|e^{i\varphi}$ such that
\begin{subequations}
\label{DMP}
\begin{align}
\varphi &  \approx\pi/2+2\omega_{0}t, 
\label{DMPa}
\\
|\mu|  &  \approx\epsilon/2,\text{ with }\epsilon\gg1 . 
\label{DMPb}
\end{align}
\end{subequations}

We now consider the resonance regime with $\kappa=2\omega_{0}$ in which the Eq. (\ref{35a}) becomes
\begin{equation}
\dot{r}\approx\varepsilon\omega_{0}\left\vert \frac{\tilde{\alpha}_{0}-\chi\tilde{\beta}_{0}}{\chi-1}\right\vert \sin{(2\omega_{0}t)}\left[\cos{\phi_{0}^{\prime}}\sin{(2\omega_{0}t)} - \sin{\phi_{0}^{\prime}}\cos{(2\omega_{0}t)}\right] ,
\label{36}
\end{equation}
whose solution is given by
\begin{equation}
r\approx r_{0}+\frac{\varepsilon}{8}\left\vert \frac{\tilde{\alpha}_{0}-\chi\tilde{\beta}_{0}}{\chi-1}\right\vert \left\{  \cos{\phi_{0}^{\prime}}\left[  4\omega_{0}t-\sin{(4\omega_{0}t)}\right]  -\sin{\phi_{0}^{\prime}}\left[1-\cos{(4\omega_{0}t)}\right]\right\}  . 
\label{37}
\end{equation}
To maximize the parameter $r$, we consider $\phi_{0}^{\prime}=0$ resulting in

\begin{equation}
r\approx r_{0}+\frac{\varepsilon}{8}\left\vert \frac{\tilde{\alpha}_{0}-\chi\tilde{\beta}_{0}}{\chi-1}\right\vert \left[  4\omega_{0}t-\sin{(4\omega_{0}t)}\right]  , 
\label{38}
\end{equation}
which for long time intervals it can be written as
\begin{equation}
r\approx r_{0}+\mathfrak{R}\frac{\varepsilon\omega_{0}t}{2} , 
\label{39}
\end{equation}
with
\begin{equation}
\mathfrak{R}=\left\vert \frac{\tilde{\alpha}_{0}-\chi\tilde{\beta}_{0}}{\chi-1}\right\vert , 
\label{R}
\end{equation}
defined in terms of the Dyson map parameters. 
It reduces to unity for the Hermitian case which corresponds to $\tilde{\alpha}_{0}=\tilde{\beta}_{0}=1$, hence Eq. (\ref{39}) retrieves the well-known result $r\approx r_{0}+\varepsilon\omega_{0}t/2$ .

\subsection{Casimir's Photon creation}

As given by Eq. (\ref{Nf}), the mean photon number generated by the DCE depends on the parameter $r$ as $\mathcal{N}(t)=\sinh^{2}{r(t)}$. 
In order to amplify this number in the pseudo-Hermitian scenario, we must construct the Dyson map to obtain $\mathfrak{R}\gg1$ which occurs when $\chi\approx1$ exactly as in Eq. (\ref{31}). Therefore, the Dyson map parameters in Eq. (\ref{DMP}) lead exactly to $\chi\approx1$ as required for the amplification in the rate of photons creation.

\begin{figure}[h!]
	\centering
	\includegraphics[width=0.51\textwidth]{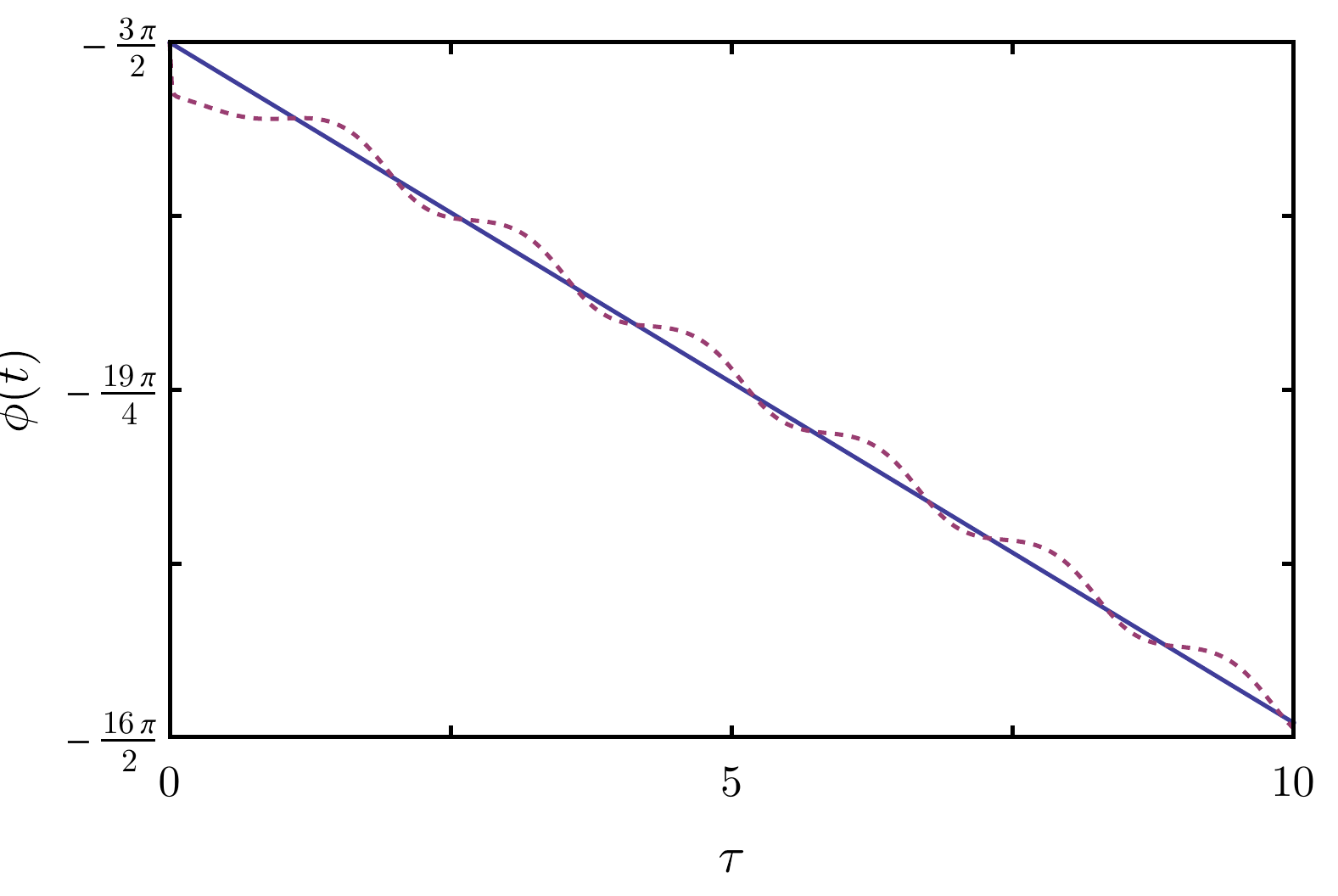}
	\caption{Plot of $\phi(t)$ against $\tau=\omega_{0}t$, considering $\omega_{0}=1$, $\tilde{\alpha}_{0}=\varepsilon=10\tilde{\beta}_{0}=10^{-2}$, $\chi=-2\Phi-1=1.0002$, $\varphi_{0}=\pi/2$, and an initial vacuum state with $r_{0}=0$ and $\phi_{0}=-3\pi/2$. The solid and dashed lines follows by	computing $\phi(t)$ analytically from Eq. (\ref{phi}) and numerically from Eqs. (\ref{24}).}
	\label{fig1}
\end{figure}

\begin{figure}[h!]
	\centering
	\includegraphics[width=0.51\textwidth]{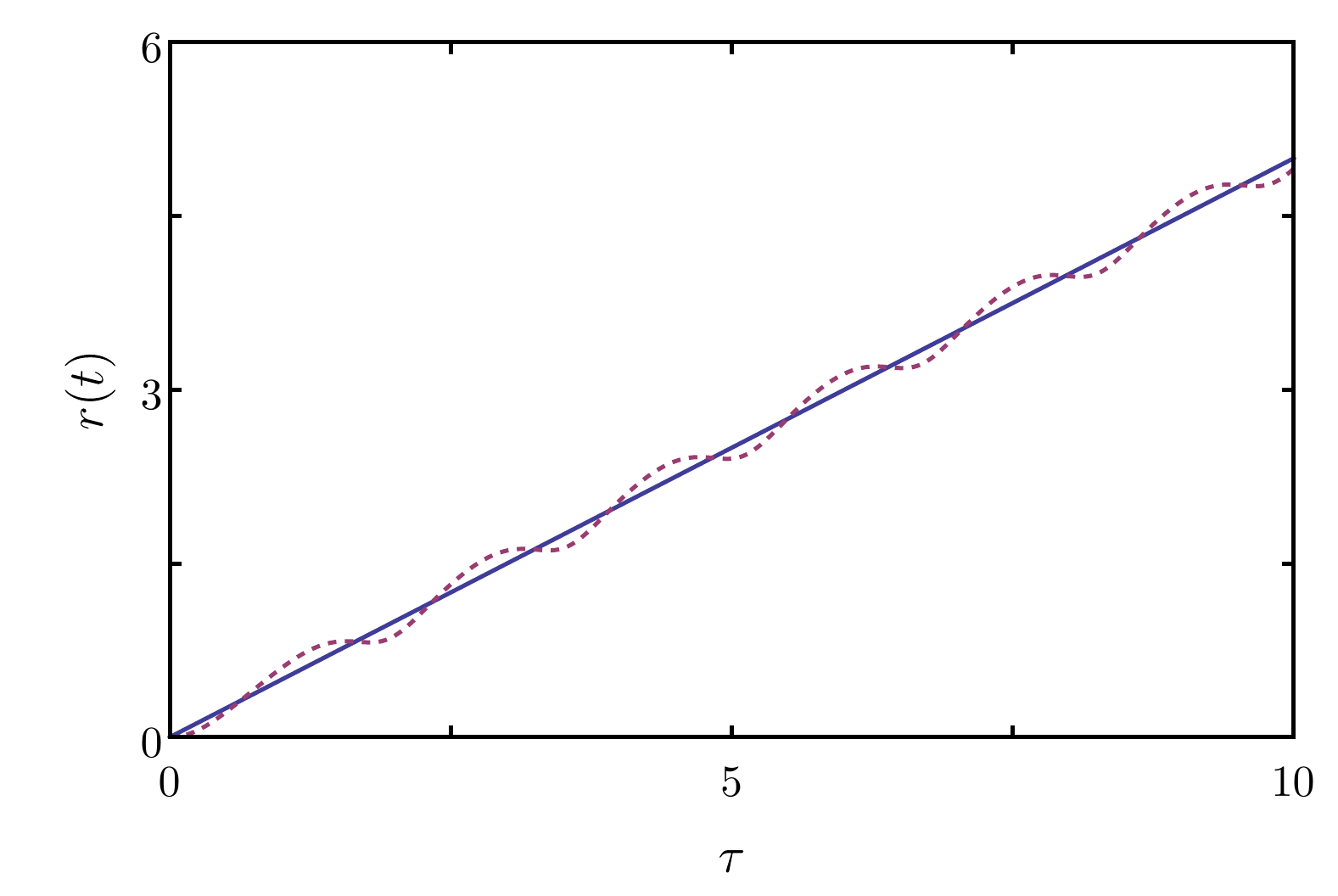}
	\caption{Plot of $r(t)$ against $\tau$, considering the same parameters used in	Fig. \ref{fig1}. The solid and dashed lines follows by computing $r(t)$ analytically	from Eq. (\ref{39}) and numerically computed from Eqs. (\ref{24}).}
	\label{fig2}
\end{figure}

In what follows we compare the solutions obtained for the squeeze parameters with those computed numerically from Eqs. in (\ref{24}). By considering $\omega_{0}=1$, $\tilde{\alpha}_{0}=\varepsilon=10\tilde{\beta
}_{0}=10^{-2}$, $\chi=-2\Phi-1=1.0002$, $\varphi_{0}=\pi/2$, and an initial vacuum state with $r_{0}=0$ and $\phi_{0}=-3\pi/2$, in Figs. \ref{fig1} and \ref{fig2} we plot the parameters $\phi(t)$ and $r(t)$, respectively, against the dimensionless time $\tau=\omega_{0}t$. In Fig. \ref{fig1}, we consider $\phi(t)$ as analytically derived in Eq. (\ref{phi}) and numerically computed from Eqs. in (\ref{24}), given by solid and dashed lines, respectively. We verify that both curves are approximately equal to each other. In Fig. \ref{fig2}, we plot the squeeze parameter $r(t)$ as derived in Eq. (\ref{39}) and numerically computed from Eq. in (\ref{24}), given by solid and dashed lines, respectively. We again verify a very good agreement between both curves, validating the approximation we made to solve the system (\ref{24}), by considering $r>1$ such that $\coth{2r\approx1}$.

\begin{figure}[htp!]
	\centering
	\includegraphics[width=0.51\textwidth]{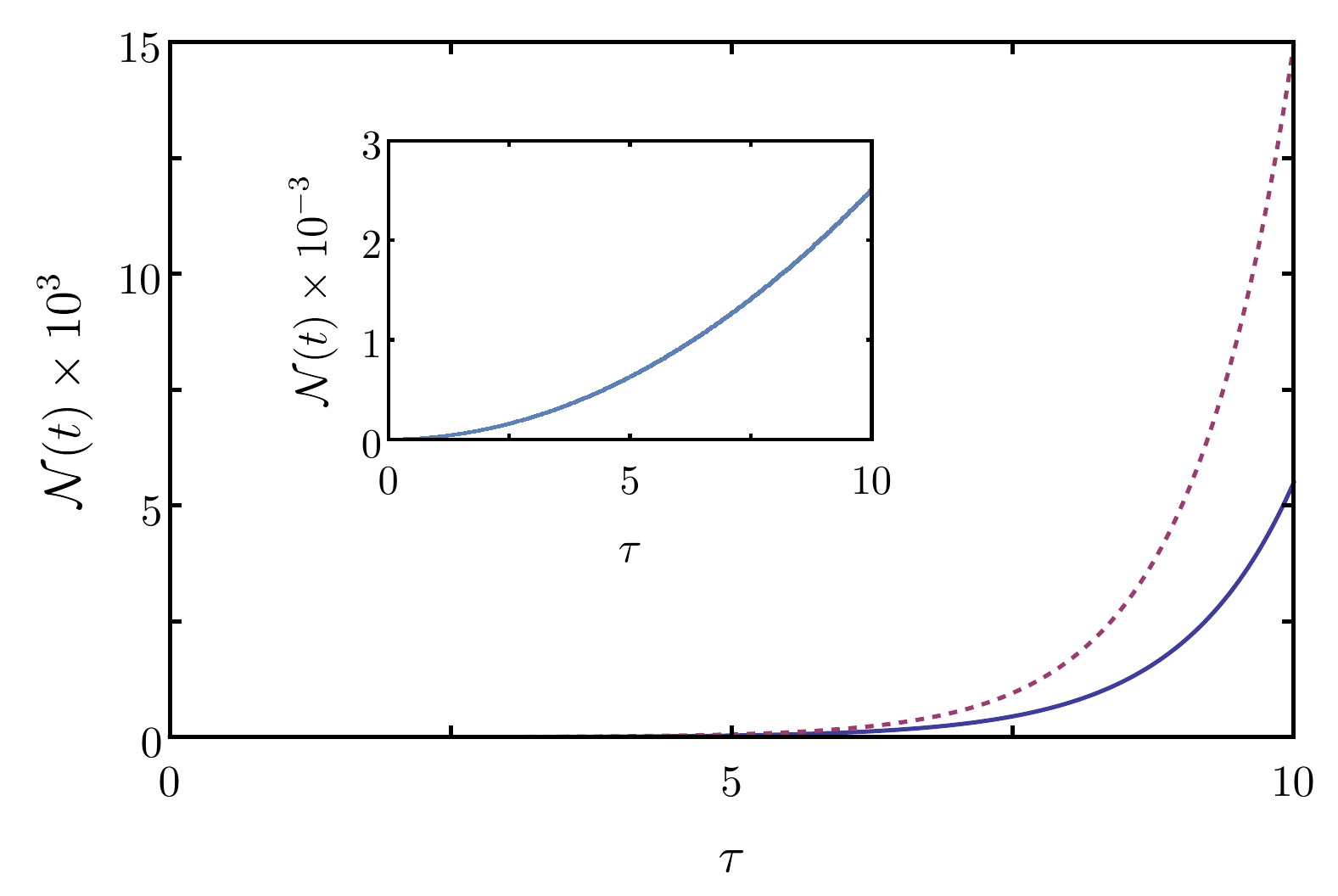}
	\caption{Plot of the mean number of photon creation $\mathcal{N}(t)$ against $\tau$, considering the same parameters used in Figs. 1 and 2, with $\tilde{\beta}_{0}=10^{-3}$ and  $\tilde{\beta}_{0}=10^{-4}$, corresponding to the solid and dotted lines, respectively. The inset plot corresponds to the Hermitian case ($\tilde{\alpha}_{0}=\tilde{\beta}_{0}=1$).}
	\label{fig3}
\end{figure}

In Fig. \ref{fig3}, we plot the mean photon number $\mathcal{N}(t)$ against $\tau$, with the same parameters considered in Figs. \ref{fig1} and \ref{fig2}, with $\tilde{\beta}_{0}=10^{-3}$ (solid line) and also $\tilde{\beta}_{0}=10^{-4}$ (dotted line) distancing the Hamiltonian (\ref{2}) even further from Hermiticity. 
It means that the rate of photon creation is higher the farther the Hamiltonian is from Hermiticity. 
In the inset, we plot the rate of photon creation for the Hermitian case,  around $6$ orders of magnitude smaller that for $\tilde{\beta}_{0}=10^{-3}$.

\section{Discussion and Conclusions}

We have considered the dynamical Casimir effect for a pseudo-Hermitian Hamiltonian, given by Eq. (\ref{2}), where an unbalanced time-dependent parametric amplification is considered apart from the usual oscillating frequency. We have verified that such an unbalanced amplification leads to a significant increase in the rate of Casimir's photon creation, in a similar way to the increment in the squeezing rate of the radiation field computed in Refs. \cite{deponte:19,dourado:20,dourado:21}. 
As remarked in Ref. \cite{deponte:19,dourado:20,dourado:21}, which demonstrate the possibility of reaching an infinite degree of squeezing at a finite time interval, \textit{pseudo-Hermitian Hamiltonians enable much higher pumping rates than those of Hermitian processes, high enough to allow infinite energy transfer in finite time. }Moreover, the farther these Hamiltonians are from Hermiticity, the higher their pumping rates.

The results following from the present work together with those in Refs. \cite{deponte:19,dourado:20,dourado:21}, pose the challenge of engineering pseudo-Hermitian effective Hamiltonians, capable of producing the higher pumping rates
necessary for high sensitive interferometry and the observation of the DCE, among many other applications. 
However, the construction of these Hamiltonians requires more energy the further away from Hermiticity we want these effective operators, what prevent us from using the method of the adiabatic elimination for the engineering protocol. 
The method of adiabatic elimination requires too many constraints between the parameters involved \cite{gamel:10,james:07}; in particular, the pumping strengths must be much smaller than the detuning between the pumping field and the cavity mode frequencies, which evidently compromises the energy supply for the construction of an effective Hamiltonian away from Hermiticity. 
We note, however, that we believe in the possibility of using the adiabatic elimination technique for the engineering of effective Hamiltonians not far from Hermiticity.

As anticipated in the Introduction, we note that rather than simulating a pseudo-Hermitian DCE through the engineering of an effective interaction, the Hamiltonian (\ref{2}) can somehow emulate a real non-Hermitian DCE. We have no clues on how to implement this last possibility.

From what is said in the above paragraphs, we conclude that the cost of achieving a strong rate of created photons by fixing $\chi$ around the unit, is a strong pumping energy for engineering the pseudo-Hermitian Hamiltonian (\ref{2}). The closer to unit is $\chi$ the stronger the energy pumping which will be automatically reverse to photon creation.

Along with the engineering of non-Hermitian Hamiltonians, another sensitive issue in our protocol for increasing the rate of Casimir photon creation is the engineering of the Dyson map. 
The experimentalist must engineer this nonunitary operator to prepare the initial state of the system described by the pseudo-Hermitian Hamiltonian (\ref{2}), i.e., $\left\vert \Psi (0)\right\rangle =\eta^{-1}(0)\left\vert \psi(0)\right\rangle $, as derived in Sec. II, $\left\vert \psi(0)\right\rangle $ being the initial state of the system described by the Hamiltonian (\ref{26}), the Hermitian counterpart of
(\ref{2}). 
The engineering of the Dyson map was addressed in Ref. \cite{dourado:21}. 
Basically, considering the Dyson map in the exponential form $\eta=e^{\mathcal{H}}$, we must engineer the operator $\mathcal{H}=\left[\tilde{\epsilon}(a^{\dagger}a+1/2)+\tilde{\mu}a^{2}+\tilde{\mu}^{\ast}(a^{\dagger})^{2}\right]  \otimes\sigma_{y}^{\mathrm{aux}}$, with $\sigma_{y}^{\mathrm{aux}}$ being an auxiliary two-level system, $\left\{  \left\vert +\right\rangle, \left\vert -\right\rangle \right\}  $, prepared in its ground state $\left\vert -\right\rangle $. 
After the evolution of the system for a time interval $\tau$ we obtain
\begin{align}
e^{-i\mathcal{H}\tau}\left\vert \psi(0)\right\rangle \left\vert - \right\rangle  
& = \exp\left\{  -\left[  \tilde{\epsilon}\tau(a^{\dagger}a+1/2)+\tilde{\mu}\tau a^{2} + \tilde{\mu}^{\ast}\tau(a^{\dagger})^{2}\right]\right\}  
\left\vert \psi(0)\right\rangle \left\vert +\right\rangle
\nonumber\\
& = \eta^{-1}(0)\left\vert \psi(0)\right\rangle \otimes\left\vert + \right\rangle , 
\label{40}
\end{align}
with $\epsilon(0)=\tilde{\epsilon}\tau$ and $\mu(0)=\tilde{\mu}\tau$, obeying Eq. (\ref{DMP}). 
After the measurement of the auxiliary system in its excited state, we thus obtain the desired prepared state $\left\vert \Psi(0)\right\rangle =\eta^{-1}(0)\left\vert \psi(0)\right\rangle $.

The Dyson map is also used for computing the observables associated with the pseudo-Hermitian system, given by Eq. (\ref{6}): $O(t)=\eta^{-1}(t)o(t)\eta(t)$. 
However, from the point of view of practical application, we do not have to worry about the time evolution of the Dyson map. There is no need to emulate the time-evolution of the Dyson map in the laboratory.
As also observed in Ref. \cite{dourado:21}, the observables $O(t)$ generally comprise linear combinations of $o(t)$, the observables associated with the hermitized Hamiltonian; consequently, the measurement of the mean values $\left\langle O(t)\right\rangle $, as defined by Eq. (\ref{ME}), requires the measurements of canonically conjugate variables, a subject addressed in Refs. \cite{wodkiewicz:84,leonhardt:93}.

Although we initially considered a Dyson map in which the modulus $\left\vert z(t)\right\vert $ and phase $\varphi(t)$ of the free parameter $z(t)=$ $\left\vert z(t)\right\vert e^{i\varphi(t)}$ are both TD functions, we later fixed $|z|\approx1$ (below unit), leaving $\varphi(t)$ still dependent on time. This choice of a time independent $\left\vert z\right\vert $ makes our analysis much simpler than that in Refs. \cite{deponte:19,dourado:20,dourado:21}, where we had to make sure that a TD $\left\vert z(t)\right\vert $ must always be limited since $|z|\in[0,1]$. 
Here, therefore, we don't have to worry about this constraint imposed on $|z|$, and the other parameters defining the Dyson map, given by Eq. (\ref{DMP}): $\varphi\approx\pi/2 + 2\omega_{0}t$ and $|\mu|\approx\epsilon/2$, with $\epsilon\gg1$, do not seem to pose difficulties for the engineering of the map via adiabatic elimination.

Finally, we note that the experimental observation of the pseudo-Hermitian DCE depends not only on the engineering of the pseudo-Hermitian Hamiltonian (\ref{2}), but on its engineering far enough away from Hermiticity, so that a measurable rate of Casimir's photon creation can be produced.

\section*{Acknowledgments}

This work was partially supported by Conselho Nacional de Desenvolvimento Cient\'{\i}fico e Tecnol\'{o}gico (CNPq), Coordena\c{c}\~{a}o de Aperfei\c{c}oamento de Pessoal de N\'{\i}vel Superior (CAPES, Finance Code
001), and Instituto Nacional de Ci\^{e}ncia e Tecnologia de Informa\c{c}\~{a}o Qu\^{a}ntica (INCT-IQ). FMA acknowledges CNPq Grants 434134/2018-0 and 314594/2020-5, and MHYM acknowledges CNPq grant 303012/2020-0.

\bibliographystyle{model1a-num-names}
\bibliography{references.bib}

\end{document}